\documentclass[twocolumn,aps,prd,   
               preprintnumbers,numbers,sort&compress,
               nofootinbib,
                            showpacs,
               colorlinks,
               linkcolor=blue,
               citecolor=blue]{revtex4-1} 

\usepackage{graphicx}
\usepackage{dcolumn}
\usepackage{bm}
\usepackage{hyperref}

\usepackage{graphicx}
\usepackage{dcolumn}
\usepackage{bm}
\usepackage{hyperref}

\usepackage{amsmath}
\usepackage[caption=false]{subfig}
\usepackage{gensymb}

\def\ra{\rangle}
\def\la{\langle}
\def\rmd{\mathrm{d}}  

\newcommand{\be}{\begin{eqnarray}}
\newcommand{\ee}{\end{eqnarray}}
\newcommand{\beq}{\begin{equation}}
\newcommand{\eeq}{\end{equation}}
\newcommand{\kmps}{\,\mathrm{km}\,\mathrm{s}^{-1}} 
\newcommand{\exclude}[1]{}
\newcommand{\Red}[1]{{\color{red}({#1})}}

\newcommand{\jcap}{JCAP}
\exclude{
 \newcommand{\prl}{Phys. Rev. Lett.}
\newcommand{\apj}{Astrophys.\ J.}
\newcommand{\nat}{Nature}
\newcommand{\prd}{Phys. Rev. D}

}
\graphicspath{{./Figures/}} 

\begin{document}
       \title{The ANITA Anomalous Events   and    Axion Quark Nuggets }
       \author{Xunyu Liang and  Ariel  Zhitnitsky}
       \affiliation{Department of Physics and Astronomy, University of British Columbia, Vancouver, V6T 1Z1, BC, Canada}
     
      \begin{abstract}
    
    
    The Antarctic Impulse Transient Antenna (\textsc{ANITA})  collaboration  \cite{Gorham:2016zah,Gorham:2018ydl,Gorham:2020zne} has reported two anomalous events with noninverted polarity. These events are hard to explain in terms of conventional cosmic rays (CRs).
    We explore a new possible explanation for these anomalous events by suggesting that these events can be related to the dark matter (DM) annihilations within the so-called axion quark nugget (AQN) DM model.  This model was initially invented for a completely different purpose to explain the observed  similarity between the dark  and the visible components  in the Universe, i.e. $\Omega_{\rm DM}\sim \Omega_{\rm visible}$ without any fitting parameters.  
    We investigate the signal properties of the upward-going AQN events, including the event rate, the pulse duration, and the electric field strength, and find them consistent with the observations.
    We list several features  of the upward-going AQN events distinct  from conventional CR air showers. The observations (or nonobservation) of these features  may  substantiate  (or refute)  our proposal. 



        \end{abstract}
     
       \maketitle

     \maketitle
\section{Introduction}\label{sec:introduction}

The  ANITA   collaboration has observed  \cite{Gorham:2016zah,Gorham:2018ydl,Gorham:2020zne} two anomalous events   that appear to be energetic cosmic showers emerging from the Earth with large exit angles. We advocate a possible explanation for these ANITA anomalous events (AAEs) in the so-called axion quark nugget (AQN) model \cite{Zhitnitsky:2002qa}. This model was invented long ago with no relation to the ANITA observations. Instead, it was dedicated to explaining the observed similarity between the dark matter (DM) and the visible densities in the Universe, i.e. $\Omega_{\rm DM}\sim \Omega_{\rm visible}$, without fine tuning. Nevertheless, we will show that an upward-going AQN event can reproduce consistent signal properties of the AAEs, including the event rate, the pulse duration, and the electric field strength.

The ANITA experiment has completed four flights \cite{Gorham:2016zah,Gorham:2018ydl,Gorham:2020zne} and reported two AAEs that are compatible with a $\nu_\tau$ neutrino interpretation of energy $\sim\,$EeV at exit angles of $-27\degree$ and $-35\degree$ relative to horizontal in the first \cite{Gorham:2016zah} and third \cite{Gorham:2018ydl} flights respectively. The radio pulse of an AAE observed by the ANITA balloon payload at an altitude of $\sim{35}\,$km is of order $(0.1-1)\,$mV/m in electric field strength and  $(1-10)\,$ns in time duration. The observed frequency spectrum of the signal is in the range $(40-800)\,$MHz. It attenuates sharply beyond the critical frequency near $800\,$MHz. Numerous suggestions \cite{Fox:2018syq,Romero-Wolf:2018zxt,Aartsen:2020vir,Motloch:2016yic,deVries:2019gzs,Dasgupta:2018dzp,Shoemaker:2019xlt,Smith:2020ecb,Chauhan:2018lnq,Heurtier:2019git,Hooper:2019ytr,Anchordoqui:2019utb,Anchordoqui:2018ucj,Dudas:2018npp,Huang:2018als,Connolly:2018ewv,Cherry:2018rxj,Anchordoqui:2018ssd,Collins:2018jpg,Chauhan:2018lnq,Heurtier:2019git,Hooper:2019ytr,Heurtier:2019rkz,Chipman:2019vjm,Abdullah:2019ofw,Anchordoqui:2019utb,Borah:2019ciw,Esmaili:2019pcy,Esteban:2019hcm,Altmannshofer:2020axr,Cline:2019snp} are proposed to explain the AAEs, but many of them suffer from difficulties; see below.

The AAEs are in critical tension with the standard model (SM) because neutrinos are exceedingly unlikely to traverse through Earth at a distance of $\gtrsim5\times10^3\,$km with such ultrahigh energy, even accounting for the $\nu_\tau$ regeneration \cite{Gorham:2016zah}. The analysis \cite{Fox:2018syq} reviewed the high-energy neutrino events from the IceCube Neutrino Observatory and inferred  that the $\nu_\tau$ interpretation is excluded by at least $5\sigma$ confidence. A similar study in Ref. \cite{Romero-Wolf:2018zxt} estimated ANITA acceptance to a $\nu_\tau$ flux and concluded at least two orders of magnitude above the upper limit from Pierre Auger Observatory and IceCube. More recently, IceCube also published severe constraints of astrophysical explanation for the AAEs under SM assumptions \cite{Aartsen:2020vir}. 
 
Alternative   explanations such as transition radiation \cite{Motloch:2016yic,deVries:2019gzs} remain unconfirmed, special reflection on a spherical surface is disfavoured \cite{Dasgupta:2018dzp}, and glaciological explanation based on subsurface reflectors \cite{Shoemaker:2019xlt} is largely excluded by ANITA \cite{Smith:2020ecb}. Notably, several beyond-the-SM (BSM) explanations are proposed \cite{Anchordoqui:2018ucj,Dudas:2018npp,Huang:2018als,Connolly:2018ewv,Cherry:2018rxj,Anchordoqui:2018ssd,Collins:2018jpg,Chauhan:2018lnq,Heurtier:2019git,Hooper:2019ytr,Heurtier:2019rkz,Chipman:2019vjm,Abdullah:2019ofw,Anchordoqui:2019utb,Borah:2019ciw,Esmaili:2019pcy,Esteban:2019hcm,Altmannshofer:2020axr}. In most cases, it suggests an origin of a (or a group of) massive hypothetical particle(s), which is strongly constrained by the IceCube and Auger bounds \cite{Cline:2019snp}. The other common problem is that the models are largely fine-tuned to match the observation of the AAEs and lack a  natural motivation.

The AQN model, as an explanation of the AAEs, does not encounter the above difficulties. The AQN is a macroscopic DM candidate with a mass of grams and a size of $0.1\rm{\,\mu m}$. Hence, an upward-going AQN event is distinct from the conventional CR air showers induced by ultrahigh-energy particles. The constraints from the Pierre Auger Observatory and IceCube are not applicable in the AQN scenario because numerous basic assumptions of the CR features are invalid.

The \textit{fundamental} parameter of the model, such as the AQN's size $R$ and the average baryon charge $\langle B\rangle$ of the AQNs, was constrained by various phenomena and observations long ago irrelevant to the AAEs. 
 In this sense, the AQN model is   rigid and predictive   since  there is little flexibility or freedom to modify the fundamental parameters mentioned above.  
This comment applies to the model itself, not to the interactions with environment, which could be complex.
 The description of the interaction  with surrounding material requires introduction of unknown \textit{phenomenological} parameters, which cannot be computed from the first principles.

In the following Sec. \ref{AQN}, we presented a short overview of the AQN model. In particular, subsection \ref{basics} introduces the fundamental facts of the model, and subsections \ref{earth} and \ref{upward} summarize the essential features relevant to the present work. Sec. \ref{proposal} describes the properties of an AQN at the instant when it emerges from the Earth's surface. Secs. \ref{sec:Expected number of events} and \ref{sec:Radio signals induced by AQN} examine the signal properties of an upward-going AQN event, including the event rate, the pulse duration, and the electric field strength. We conclude in Sec. \ref{conclusion}, where we explicitly formulate some distinct features between upward-going AQN events and conventional CR events. We also suggest possible tests which may support or refute our proposal.


 \section{The AQN   DM model }\label{AQN}
 
 

We start with a few historical remarks and motivation of the AQN model in subsection \ref{basics}. In subsection \ref{earth}, we overview several recent observations possibly related to an AQN hitting the Earth, including the puzzling bursts observed by the Telescope Array experiment, the exotic events recorded by the Pierre Auger observatory, and the Multi-Modal clustering events observed by HORIZON 10T instrument.
In subsection \ref{upward}, we review specific characteristics of the AQNs traversing the Earth, such as internal temperature and ionization level. These characteristics will be necessary for interpreting the AAEs as the Earth-emergent AQN events.
 
 
 \subsection{The basics 
 }\label{basics}
 

 
 

The AQN DM model [31] was solely motivated to explain the observed similarity between the dark and the visible matter densities, i.e. $\Omega_{\rm DM}\sim \Omega_{\rm visible}$, in the Universe without fine tuning. The AQN is in many respects similar to Witten's quark nuggets (see  \cite{Witten:1984rs,Farhi:1984qu,DeRujula:1984axn}  and review \cite{Madsen:1998uh}). This type of DM candidate is ``cosmologically dark" because of its small cross-section-to-mass ratio, not the weakness of its interactions. Namely, all quark nugget models have an excessively diffuse number density that obscures many observable consequences despite their strongly-interacting nature. We refer to the original papers \cite{Liang:2016tqc,Ge:2017ttc,Ge:2017idw,Ge:2019voa} devoted to the AQN formation, including the generation of the baryon asymmetry and the survival pattern, in the early Universe with its unfriendly environment; see also a recent brief review article \cite{Zhitnitsky:2021iwg} for many subtle questions on the formation mechanism. Here we mention several essential points to benefit the readers and make the presentation self-contained.

Compared to the original quark nugget models \cite{Witten:1984rs,Farhi:1984qu,DeRujula:1984axn,Madsen:1998uh}, the new element in the AQN model is the presence of the axion domain walls, which are copiously produced during the QCD transition\footnote{
The axion is arguably the most compelling solution to the so-called strong $\cal{CP}$ problem (see original papers on the axion \cite{axion1,axion2,axion3,KSVZ1,KSVZ2,DFSZ1,DFSZ2} and   recent reviews \cite{vanBibber:2006rb, Asztalos:2006kz,Sikivie:2008,Raffelt:2006cw,Sikivie:2009fv,Rosenberg:2015kxa,Marsh:2015xka,Graham:2015ouw, Irastorza:2018dyq}). As we will discuss, a globally coherent axion field is responsible for the baryon asymmetry in the early Universe. However, this source of $\cal{CP}$ violation is no longer available at the present time due to the axion dynamics.}. The domain wall plays a dual role. First, it serves as an additional stabilization factor for the nuggets, which helps alleviate many problems with the original nugget construction \cite{Witten:1984rs,Farhi:1984qu,DeRujula:1984axn,Madsen:1998uh}.  Secondly, the same axion field $\theta (x)$ generates the strong and coherent $\cal{CP}$ violation in the entire visible Universe. 

The inherent assumption in the AQN model is that the Peccei–Quinn (PQ) phase transition happens before inflation. Consequently, the initial misalignment angle $\theta_0\neq 0$ assumes one and the same value over the enormous scale of the visible Universe\footnote{One should comment here that our scenario is dramatically different from conventional studies of the topological defects  when PQ phase transition happens after inflation. In contrast, we assume the PQ phase transition happens before inflation, in which case the axion strings are not present in the system. However, the $N_{\rm DW}=1$ domain walls can be formed.}. The axion field $\theta (x)$ can be treated as a classical $\cal{CP}$ violating field correlated on the scale of the entire Universe before the QCD epoch. The axion field starts to oscillate at the QCD transition by emitting the propagating axions. However, these oscillations remain coherent on the scale of the entire Universe. Therefore, the $\cal{CP}$ violating phase remains coherent on a global scale.

One related conceptual question would be: why the baryons do not easily leak through the wide and flat axion domain? The answer is that the domain wall not only comprises axions but also mixes with the singlet $\eta'$ field. This $\eta'$ substructure has a narrow width of $\Lambda_{\rm QCD}^{-1}$ and interacts strongly with baryons. At the same time, it contributes negligibly to the surface tension of the domain wall. This coupling is a very generic feature of the system because the axion and $\eta'$ fields are the phases of the same chiral condensate, and they interpolate between two physically identical (but topologically distinct) states  as a combination of two knotted fields. 
The mixing also implies metastability of the domain walls, as the $\eta'$ and the axion fields cannot be unknotted separately. Consequently, the axion domain wall (with the QCD substructure) is topologically stable even if it has a width (order of $ m_a^{-1}\sim\,$cm)  much larger than the size of the quark nugget (order of $0.1\,\mu$m). To summarize, the presence of the $\eta'$ substructure stabilizes the domain wall bubbles and prevents the baryon charge leakage.



 




Another conceptual question would be the evolution of these closed bubbles in the cosmic plasma. This complex dynamic problem involves immense differences in scale, such as the QCD scale $\Lambda_{\rm QCD}$, the axion mass $m_a$, and the cosmic time $\sim10^{-4}\,$s at the QCD epoch. First, collapses of the closed $N_{\rm DW}=1$ bubbles will be halted by the Fermi pressure from the accumulated fermions. Eventually, the $N_{\rm DW}=1$ domain wall bubbles will become stable AQNs and comprise the dark sector. Furthermore, the chemical potential inside the bubbles assumes a sufficiently large value $\mu_{\rm form}\gtrsim 400\,$MeV during this long evolution, see the orange line in Fig. 2 in \cite{Ge:2019voa}. The magnitude of $\mu_{\rm form}$ supports the AQN formation in the colour-superconducting (CS) phase. As mentioned above, the corresponding evolution is rather complex as it includes three immensely different scales. Nevertheless, it leads to a consistent and coherent picture. 
 


 






The other new element of the  AQN model, which plays
an absolutely crucial role for  the present work,  is that nuggets can be made of {\it matter} as well as {\it antimatter} during the QCD transition. 
Because of the coherent $\cal{CP}$ violation in the entire Universe, there is a preferential production of one species of the AQNs (i.e. the antimatter AQNs) over the other (i.e. the matter AQNs). The preference is determined by the sign of the initial misalignment angle $\theta_0$ at the beginning of the AQN formation. Consequently, the dark and the visible matter densities will automatically assume the same order of magnitude, i.e. $\Omega_{\rm DM}\sim \Omega_{\rm visible}$, without fine tuning.


For the present studies, however,  we take the  agnostic viewpoint, and assume that such antimatter AQNs 
are present in our Universe today irrespective to their   formation mechanism. This assumption is consistent with all presently available cosmological, astrophysical and terrestrial  constraints as long as  the average baryon charge of the nuggets is sufficiently large as we review  below.

One should emphasize that AQNs are absolutely stable configurations on cosmological scales. Furthermore, the antimatter which is hidden  in form of the very dense nuggets is unavailable for annihilation unless the AQNs hit the stars, the planets, or interstellar medium. 

However, when the AQNs hit  the stars, the planets, or interstellar medium it may lead to observable phenomena. In particular,  the injection of the energy due to the AQNs hitting  the Sun 
  may explain\footnote{In fact, to resolve this problem  Parker conjectured  long ago  \cite{Parker} that ``nanoflares"   are   identified with the  annihilation events in the AQN framework.
    The luminosity  of the Extreme UV (EUV) radiation from corona due to these annihilation events is unambiguously determined by the DM density. It is very nontrivial consistency check that  the computed  luminosity from the corona nicely matches with observed EUV radiation.}  the ``Solar Corona heating problem" as advocated in \cite{Zhitnitsky:2017rop,Raza:2018gpb,Ge:2020xvf}.   
 There are also very rare events of annihilation in   the galaxy, which, in fact, may explain some observed galactic excess emissions in different frequency bands,
 including very mysterious diffuse  UV radiation   \cite{Henry_2014,Akshaya_2018} as recently argued in  \cite{Zhitnitsky:2021wjb}. 
 
 The strongest direct detection limit\footnote{Non-detection of etching tracks in ancient mica gives another indirect constraint on the flux of   DM nuggets with mass $M> 55$g   \cite{Jacobs:2014yca}. This constraint is based on assumption that all nuggets have the same mass, which is not the case  as we discuss below.
The nuggets with small masses represent a tiny portion of all nuggets in this model, such that this constraint is easily  satisfied with any reasonable nugget's size distribution.} is  set by the IceCube Observatory's,  see Appendix A in \cite{Lawson:2019cvy}:
\be
\label{direct}
\la B \ra > 3\cdot 10^{24} ~~~[{\rm direct ~ (non)detection ~constraint]}.
\ee
Similar limits are   also obtainable 
from the   ANITA 
  and from  geothermal constraints which are also consistent with (\ref{direct}) as estimated in \cite{Gorham:2012hy}. It has been also argued in \cite{Gorham:2015rfa} that that AQNs producing a significant neutrino flux 
in the 20-50 MeV range cannot account for more than 20$\%$ of the DM 
density. However, the estimates \cite{Gorham:2015rfa} were based on assumption that the neutrino spectrum is similar to  the one which is observed in 
conventional baryon-antibaryon annihilation events which typically produce a large number of pions and  muons and thus generate a 
significant number of neutrinos and antineutrinos in the 20-50 MeV range where 
SuperK has a high sensitivity. However, the critical difference in the case of AQNs  is that the annihilation proceeds within the 
colour superconducting (CS) phase where the energetics are drastically  different \cite{Lawson:2015cla}. The main point is that, in most CS phases, the lightest pseudo Goldstone mesons 
(the pions and kaons) have masses in the  20 MeV range, 
rather than 140 MeV in hadronic phase. This dramatically changes 
entire spectrum such that the  main assumption of \cite{Gorham:2015rfa} on similarity of the neutrino's spectrum in both phases is incorrect. The  resulting flux computed in \cite{Lawson:2015cla} is perfectly consistent with observations. Furthermore, precisely these low energy ($\lesssim 20~ \rm MeV$) AQN-induced neutrinos produced in the Earth's interior might be responsible  for explanation of the long standing puzzle of the DAMA/LIBRA  observation  of the annual modulation at $9.5\sigma$ confidence level as argued in \cite{Zhitnitsky:2019tbh}. 

The authors of Ref. \cite{SinghSidhu:2020cxw} considered a generic constraint for the nuggets made of antimatter (ignoring all essential  specifics of the AQN model such as quark matter  CS phase of the nugget's core). Our constraints (\ref{direct}) are consistent with their findings including the Cosmic Microwave Background (CMB), Big Bang Nucleosynthesis (BBN), and others, except the constraints derived from    the so-called ``Human Detectors". 
As explained in \cite{Ge:2020xvf}
  the corresponding estimates of Ref. \cite{SinghSidhu:2020cxw} are   oversimplified   and do not have the same status as those derived from CMB or BBN constraints\footnote{In particular, the rate of energy deposition was estimated in \cite{SinghSidhu:2020cxw} assuming that the annihilation processes between antimatter nuggets and baryons are similar to $p\bar{p}$ annihilation process. It is known that it cannot be the case because the annihilating objects have drastically different internal structures (hadronic phase versus CS phase). It has been also assumed in \cite{SinghSidhu:2020cxw} that  a typical X ray energy  is around  1 keV, which is  much lower than direct computations in the AQN model would  suggest \cite{Budker:2020mqk}.
  \exclude{ Higher energy X rays have much longer mean free path, which implies that the dominant portion of the energy will be deposited outside the human body. Finally, the authors of Ref.  \cite{SinghSidhu:2020cxw} assume that an antimatter nugget will result in ``injury   similar to a gunshot". It is obviously a wrong picture as the size of a typical nugget is only $R\sim 10^{-5} {\rm cm}$ while the most of the energy is deposited in form of the X rays on centimeter  scales \cite{Budker:2020mqk} without making a large hole similar to  bullet as assumed in \cite{SinghSidhu:2020cxw}. In this case a human's  death may occur as a result of  a large dose of radiation with a long time delay, which would make  it hard to identify the cause of the death. This argument (about the time delay and difficulties with possible death identification) should be contrasted   with the main assumption of \cite{SinghSidhu:2020cxw} that all such cases would be unambiguously and quickly identified. While in the journal version of \cite{SinghSidhu:2020cxw} the constraint had been weaken in comparison with the arXiv version, we still think it is not adequately reflect the complex physics as outlined  above.}}.

 While ground based direct searches   
offer the most unambiguous channel
for the detection of quark nuggets   
the flux of nuggets   is inversely proportional to the nugget's mass   and 
consequently even the largest available conventional DM detectors are incapable  to exclude    the entire potential mass range of the nuggets. Instead, the large area detectors which are normally designed for analysing     the high energy cosmic rays are much better suited for our studies of the AQNs as we discuss in next section \ref{earth}.

 \subsection{
 AQNs entering the Earth
 }\label{earth}
    For our present work, however,  the  most relevant studies  are related to the effects which  may occur when the antimatter AQNs hit the Earth  and continue to  propagate in deep underground in very dense environment.
   In this case,  most of the energy injection will   occur  in the Earth's interior.    The corresponding signals are very hard to detect  as the photons, the electrons and the positrons will be quickly absorbed by surrounding dense  material deep underground, while the emissions of the very weakly interacting neutrinos and axions are hard to recover. \exclude{Nevertheless, as we already mentioned,   the AQN-induced neutrinos produced in the Earth's interior might be responsible  for explanation of the long standing puzzle of the DAMA/LIBRA  observation  of the annual modulation  \cite{Zhitnitsky:2019tbh}. }Nevertheless, the AQN-induced axions from deep interior can be recovered by analyzing    the daily and annual modulations  as suggested in \cite{Budker:2019zka} and elaborated in \cite{Liang:2020mnz}.      The AQN annihilation events in the Earth's atmosphere could produce  infrasound and seismic acoustic waves     as discussed in   \cite{Budker:2020mqk,Figueroa:2021bab},  when the infrasound and seismic acoustic waves indeed have been recorded  by  dedicated instruments.
   \exclude{\footnote{A single observed event properly recorded by  the Elginfield Infrasound Array  (ELFO) which was  accompanied by correlated seismic waves is dramatically different from conventional meteor-like  events. In particular, while the event was very intense it has not been detected by a synchronized  all-sky camera network (visible frequency bands)  which ruled out a meteor source. At the same time this event is consistent  with   interpretation of the    AQN-induced    event because  the visible frequency bands must be strongly suppressed when  AQN propagates in atmosphere \cite{Budker:2020mqk} }}  Furthermore,   the AQN annihilation events explain  the recently observed puzzling  
   cosmic ray (CR) like events such as the mysterious bursts  observed  by  the Telescope Array (TA) \cite{Abbasi:2017rvx,Okuda_2019},  the exotic events observed by the Pierre Auger \cite{PierreAuger:2021int,2019EPJWC.19703003C,Colalillo:2017uC}    and the very puzzling Multi-Modal clustering  events observed by the HORIZON 10T \cite{2017EPJWC.14514001B,Beznosko:2019cI}.   In these  cases the mysterious events can be explained as  the AQN annihilation events    as argued in \cite{Zhitnitsky:2020shd,Liang:2021wjx} for the TA events, in \cite{Zhitnitsky:2022swb} for the Pierre Auger exotic events and in   \cite{Zhitnitsky:2021qhj} for the Multi-Modal Clustering  Events.

Finally,  the seasonal variations of the X ray   background in the near-Earth environment in the 2-6 keV energy range as observed by the  XMM-Newton at  $ 11\sigma$ confidence level  \citep{Fraser:2014wja} may be also naturally explained within the same AQN framework as argued \cite{Ge:2020cho}. 
This  application to the X ray emission in the near-Earth environment  is especially relevant for the present work because the AQN-induced  X rays   according to the proposal \cite{Ge:2020cho} are  originated from the AQN upward-going (Earth emergent) events when the AQNs traversed  through the Earth interior and exited the Earth's surface. 

Such events could, in principle, be responsible for the ANITA mysterious events \cite{Gorham:2016zah,Gorham:2018ydl,Gorham:2020zne}   with exit angles  of $ -27\degree$ and $ -35\degree$   relative to the horizon    as advocated in present work. Before we present our arguments  we have to highlight in next section the basic characteristics of the AQNs traversing the Earth, which is the topic of the next subsection.   
    
     \subsection{Upward-going (Earth-emergent) events}\label{upward}
     
      The goal  here is to explain the basic features of the AQNs when they enter the dense regions of the surrounding material and annihilation processes start. 
       The related  computations originally have been carried out in \cite{Forbes:2008uf}
 in application to the galactic environment with a typical density of surrounding visible baryons of order of $n_{\rm galaxy}\sim 300 ~{\rm cm^{-3}}$ in the galactic center, in dramatic contrast with dense region in the Earth's interior  when $n_{\rm rock}\sim 10^{24} ~{\rm cm^{-3}}$. We review  these computations with a few additional elements, which must be implemented in case of the propagation in the  Earth's atmosphere  and  interior when the density of the environment is much greater than in the galactic  environment.  

The total surface
emissivity from the electrosphere has been computed in \cite{Forbes:2008uf}, given by 
\begin{equation}
  \label{eq:P_t}
  F_{\text{tot}} \approx 
  \frac{16}{3}
  \frac{T^4\alpha^{5/2}}{\pi}\sqrt[4]{\frac{T}{m}}\,,
\end{equation}
 where $\alpha\approx1/137$ is the fine structure constant, $m=511{\rm\,keV}$ is the mass of electron, and $T$ is the internal temperature of the AQN.  
 One should emphasize that the emission from the electrosphere is not thermal, and the spectrum is dramatically different from blackbody radiation (see \cite{Forbes:2008uf} and also Appendix \ref{details} with more details). 
   
A typical internal temperature   of  the  AQNs for a very dilute galactic environment can be estimated from the condition that
 the radiative output of Eq. (\ref{eq:P_t}) must balance the flux of energy onto the 
AQN 
\be
\label{eq:rad_balance}
    F_{\text{tot}}  (4\pi R^2)
\approx \kappa\cdot  (\pi R^2) \cdot (2~ {\rm GeV})\cdot n \cdot v_{\rm AQN},  
\ee 
where $n$ represents the number density of the environment.  The left hand side accounts for the total energy radiation from the  AQN's surface per unit time as given by (\ref{eq:P_t}). The right hand side  accounts for the rate of annihilation events when each successful annihilation event of a single baryon charge produces $\sim 2m_pc^2\approx 2~{\rm GeV}$ energy. In Eq. \,(\ref{eq:rad_balance}), we assume that  the AQN is characterized by the geometrical cross section $\pi R^2$ when it propagates 
in an environment with a local density $n$ with velocity $v_{\rm AQN}\sim 10^{-3}c$.

The factor $\kappa$ is introduced to account for the fact that not all matter striking the  AQN will 
annihilate and not all of the energy released by an annihilation will be thermalized in the  AQNs by changing the internal temperature $T$. In particular,  some portion of the energy will be released in the form of the axions, the neutrinos and the electron-positron pairs by the mechanism  discussed below. 
  In a neutral dilute environment considered  previously \cite{Forbes:2008uf},  the value of $\kappa$ cannot exceed $\kappa \lesssim 1$, which would 
correspond to the total annihilation of all impacting matter into thermal photons. The high probability 
of reflection at the sharp quark matter surface lowers the value of $\kappa$. The propagation of an ionized (negatively charged) AQN in a  highly ionized plasma (such as solar corona)   will 
 increase  the effective cross section. As a consequence,   the value of  $\kappa$ could be very large as discussed in \cite{Raza:2018gpb} in application to the solar corona heating problem.
 
 The  internal AQN temperature  had been estimated previously for a number of  cases.  It  may assume dramatically different values, mostly due to the huge difference in the number density $n$ entering (\ref{eq:rad_balance}). In particular, for the galactic environment $T_{\rm galaxy}\approx 1$\,eV, while in the 
 deep Earth's interior it could be as high as  $T_{\rm rock}\approx  (100-200) $\,keV. Precisely this value of  $T$  had been used as the initial temperature 
of the AQNs in  the proposal \cite{Ge:2020cho} explaining the seasonal variations of the X rays observed the  XMM-Newton at  $ 11\sigma$ confidence level  \citep{Fraser:2014wja} at distances $r\sim (6-10) R_{\oplus}$ from the Earth's surface. The same temperature 
has been also used in  \cite{Zhitnitsky:2021qhj} for explanation of the Multi-Modal clustering CR-like  events  observed by the HORIZON 10T \cite{2017EPJWC.14514001B,Beznosko:2019cI}.  For our estimates in the present  work, we shall use the same
 $T_{\rm rock}\approx  (100-200) $\,keV for explaining the AAEs.
 \exclude{\footnote{\label{cooling}It is very unlikely for the temperature to reach  much higher  values    because a different cooling mechanism ($e^+e^-$ pair production) becomes very  efficient as $T$ reaches the region which is   relatively close to $m=511$ keV   such that the anticipated  suppression $\exp(-2m/T)$   becomes less dramatic for $T\gtrsim 10^2$ keV, see also Appendix \ref{details} with more details.}. 
 The difference with previous studies  \cite{Ge:2020cho} is that 
 in this work we are interested in the instant  when  the very hot AQNs   just cross the Earth surface and enter the atmosphere (upward events).
 As we argue below the radio pulse is emitted precisely at this moment.  It should be contrasted with  the   applications considered in \cite{Ge:2020cho} when the  nuggets had travelled $\sim 10^4$ km in empty space before   emitted X rays  could be  detected by XMM-Newton observatory. 
   }
  
  One more feature   we want to mention here and which is relevant for our present studies is as follows. 
  There are many 
  consequences of  the high internal temperature   $T\approx (100-200) $\,keV, which results from annihilation processes as mentioned above. First, some positrons from the electrosphere may get excited and even leave the system, which obviously results 
  in ionization of the AQN itself. Secondly,  the $e^+e^-$ pairs may be produced at  such a high temperature  as the conventional suppression factor 
  $\exp (-2m/T)$ is not dramatic  for such $T$. This  is the key element for the present work, as will be explained in next section \ref{proposal}. 

\section{ANITA anomalous  events as upward-going AQN events}\label{proposal}

We will now relate the AAEs to the upward-going AQN events with the specific features described in Sec. \ref{AQN}. We highlight the primary ideas here and refer readers to Appendix \ref{details} for technical details.

As mentioned, the AQNs propagating in the Earth's interior have a high temperature of $10^2\,$keV. It enables productions of $e^+e^-$ pairs in the AQN's electrosphere. The $e^+e^-$ pair production in a hot and dense environment has been studied previously in the context of the quark stars (see references in Appendix \ref{details}). However, the developed technique does not apply to the present work, as the thermal equilibrium is absent in a small-size system such as the AQN. Thus, we will introduce a phenomenological parameter $N$, the number of the produced $e^+e^-$ pairs at the moment of an AQN crossing the Earth's surface, in what follows.

The fates of an electron and a positron are entirely different  in the background of an intense electric field. The electric field is coming from the ionization of the hot AQN. As estimated in Appendix \ref{details}, the electric field repulses the created electrons and accelerates them to ultrarelativistic energy, $\la E\ra \sim 10{\rm\,MeV}$, as given by Eq. \eqref{E}. In contrast, the same electric field attracts the created positrons and traps them on the surface of the AQN. One should emphasize that these positrons created by pair productions have fundamentally different properties from those orbiting in the electrosphere \eqref{density}. The orbiting positrons have much smaller bound energies and can locate far away from the surface of the AQN.

Consequently, we expect an instantaneous emission of energetic electrons with typical energy $\langle E\rangle\sim10\,$MeV, when the AQN emerges from the Earth's surface due to drastic perturbation. This instantaneous emission differs from when an AQN propagates in the Earth's deep interior. First, the emitted energetic electrons can propagate for several
kilometres in the atmosphere and produce observable effects (e.g. geosynchrotron radiations). This is in contrast to when an AQN propagates underground, where the photons and the electrons will be quickly absorbed. Second, the electrons are dominantly emitted in the direction of the AQN velocity. This is because collisions and annihilations concentrate on the incident side of the AQN. As being created via pair production, the electrons are immediately accelerated along the direction of the background electric field, which approximately  coincides with the direction of the AQN velocity. We also expect that the cross section of the incident collisions given by the nugget's size $R$, which enters formula \eqref{eq:rad_balance}. It should be contrasted with a much larger size $r^*$ of the electrosphere at high temperature, as discussed in Appendix \ref{details}.
 

Any precise computation of this instant of crossing is a very hard problem  of non-equilibrium dynamics, which is beyond the scope of the present work. Fortunately, the observable radio  signal  (which is a  direct consequence   of the energetic emitted electrons) is not very sensitive to the details of this non-equilibrium mechanism and corresponding time scales as it depends on several basic parameters such as the typical energy $\la E\ra \sim 10~ {\rm  MeV}$ of the emitted electrons as estimated in (\ref{E}) and the number of emitted electrons $N$ as mentioned earlier. It is assumed that this number $N$  will  generate    a coherent radio signal. The parameter $N$ determines the intensity of the radiation, and cannot be computed from the first principles due to the very large uncertainties  of the  complicated non-equilibrium dynamics such as turbulence, shock waves, strong ionization as mentioned above and further  elaborated  in Appendix
 \ref{details}. This  parameter   must satisfy the constraint $N\ll N_{\rm max}$ where  $N_{\rm max}$ is the maximal number of potentially available electrons   (\ref{N}) which could be, in principle, liberated from the AQN at the instant of crossing the Earth's surface.    
 
  The number of electrons $N$ which are emitted by a conventional cosmic ray showers of energy   $E_{\rm CR}\sim (10^{17}-10^{18})~ \rm eV$ is of order $N\sim (10^8-10^{9})$. We anticipate   a similar magnitude for the number of electrons $N\sim (10^8-10^{9})$ emitted by AQNs  as the observed intensity of the field strength  for the anomalous ANITA events of order $\rm mV/m$ which agrees with the value of $N$ assuming the  shower energy is $E_{\rm CR}\sim (10^{17}-10^{18})~ \rm eV$.  
  
  Important arguments supporting our proposal that the AQN induced events could mimic the anomalous radio signals  observed by ANITA are  based on very specific qualitative characteristics such as the  spectrum and the pulse duration, rather than on a precise estimation of parameter $N$. The  event rate  of such anomalous events is also shown to be  consistent  with our AQN interpretation 
  (see next section \ref{sec:Expected number of events}). 
 
 Furthermore, the average  electron's  energy in the energetic CR events with $E_{\rm CR}\sim (10^{17}-10^{18})~ \rm eV$ is around 30 MeV which is in the same energy range of the AQN-induced electrons as estimated in (\ref{E}). Therefore, the electrons which are released as a result of AQN crossing the boundary in upward-going (Earth emergent) event could mimic the radio signal of the conventional CR shower events as detected by ANITA. 
 The topic of Sec.  \ref{sec:Radio signals induced by AQN} will be the estimation of specific properties of the AQN-induced radio signal, such as the spectrum, the pulse duration, and the electric field strength.
 
\section{The Event Rate of AAEs }
\label{sec:Expected number of events}
 This section is devoted to estimate the event rate of AAEs within the AQN framework. It is expected to be a qualitative estimate up to an order-of-magnitude check due to large uncertainties in parameters and rare occurrence of the observed AAEs. Nevertheless we would like to present such estimate to demonstrate that our interpretation of AAEs as a consequence of upward-going AQNs is at least a self-consistent proposal.

The expected number of the AAEs assuming that they are induced by the AQNs   can be  estimated as follows:
\begin{equation}
\label{eq:cal N}
{\cal N}
\approx {\cal{A}}_{\rm eff} {\cal T}\Delta\Omega
\frac{\rmd\Phi}{\rmd A\rmd\Omega}\,,
\end{equation}
where  ${\cal{A}}_{\rm eff}\approx4{\rm\,km}^2$ is the effective area of ANITA \cite{Gorham:2018ydl}, ${\cal T}\approx48.75\,$days is the combined exposure time of ANITA\footnote{\label{ft:exposure time}The effective exposure time was 17.25 days for ANITA-I \cite{Schoorlemmer:2015afa}, 7 days for ANITA-III \cite{Gorham:2018ydl}, and 24.5 days for ANITA-IV \cite{Gorham:2020zne}, where we exclude ANITA-II (28.5 days) as it is not sensitive to upward-going air showers.},  $\Delta\Omega\approx2\pi$ for isotropic flux of AQNs, $\Phi$ is the total hit rate of AQNs on Earth \cite{Lawson:2019cvy}:
\begin{equation}
\label{Phi}
\begin{aligned}
\Phi
&\approx2.12\times10^7{\rm\,yr}^{-1}  \\
&\quad\times\left(\frac{\rho_{\rm DM}}{0.3{\rm\,GeV\,cm^{-3}}}\right)
\left(\frac{v_{\rm AQN}}{220\kmps}\right)
\left(\frac{10^{25}}{\langle B\rangle}\right)\,,
\end{aligned}
\end{equation}
where $\rho_{\rm DM}$ is the local density of DM. The local rate of upward-going AQNs per unit area depends on the flux distribution of the AQN:
\begin{equation}
\label{Phi1}
\begin{aligned}
\frac{\rmd\Phi}{\rmd A\rmd\Omega}
=\frac{\eta}{4\pi R_\oplus^2}\Phi  =  4\cdot 10^{-2}\left(\frac{10^{25}}{\langle B\rangle}\right)\rm \frac{\eta ~events}{yr\cdot  km^2}
\end{aligned}
\end{equation}
where $R_\oplus=6371\,$km is the radius of the Earth, and $\eta$ is a parameter that characterizes the local flux distribution of AQN. Specifically, $\eta\approx1$  for isotropic distribution, and $\eta\approx2$ for the so-called fixed-wind distribution based on standard halo model so that more AQNs enter the Earth from the northern hemisphere and exit in the southern hemisphere  \cite{Liang:2019lya}. Note that the survival rate of an AQN traverses through Earth is also taken into account in $\eta$ implicitly, but it is an exceedingly minor effect comparing to the flux distribution because AQNs can penetrate through the Earth easily based on Monte Carlo simulation in Ref. \cite{Lawson:2019cvy}. 

Combining the estimates above, one should have
\begin{equation}
\label{eq:cal N ::num}
{\cal N}
\approx0.28
\left(\frac{\eta}{2}\right)
\left(\frac{\rho_{\rm DM}}{0.3{\rm\,GeV\,cm^{-3}}}\right)
\left(\frac{v_{\rm AQN}}{220\kmps}\right)
\left(\frac{10^{25}}{\langle B\rangle}\right)\,.
\end{equation}
The expected number of events ${\cal N}\approx0.3$ is    almost one order of magnitude lower than  ${\cal N}_{\rm obs}=2$ events  observed by ANITA. Nevertheless, we consider this order of magnitude estimation  (\ref{eq:cal N ::num}) being  consistent with our proposal due to many uncertainties which 
enter this estimate.

First, the parameters in Eq. \eqref{eq:cal N ::num} are in fact not precisely known. Essential parameters such as $\eta$, $\rho_{\rm DM}$, $\langle v_{\rm AQN}\rangle$, and $\langle B\rangle$ only have accuracy up to order one as the local flux distribution of DM and size distribution of AQN remain unknown to date.    In fact, there are numerous hints suggesting that   $\rho_{\rm DM}$ locally in solar system could be much larger from its canonical value, see a short comment on this with the references in the last paragraph   of this section. 
Similarly, the effective area of detection ${\cal{A}}_{\rm eff}$ may double depending on the exit angle (see estimate in e.g. Ref. \cite{Huang:2018als}), and the effective exposure time is potentially longer than our conservative estimate if we take into account the ANITA-II flight in the estimate (see footnote \ref{ft:exposure time}).    
In addition, the total number of observed AAEs could be a statistical fluctuation  due to its rare occurrence  (only 2).

We   consider this order of magnitude estimate (\ref{eq:cal N ::num}) as a highly nontrivial consistency check of our proposal  as the basic numerical factors entering  (\ref{eq:cal N ::num}) had been fixed from dramatically different physics (including solar corona heating puzzle) and  can easily deviate by large factor. More importantly, our main arguments leading to the  identification of the AAE with the AQN induced radio pulses
are based on specific qualitative features such as frequency dependence and duration of the pulse which are not sensitive   to these huge uncertainties  in the normalization factor (\ref{eq:cal N ::num}).  We consider the agreement between the observations and our theoretical estimates (to be discussed in next section \ref{sec:Radio signals induced by AQN})  for these specific  characteristics  as the  strong  arguments supporting our identification.  

It is also interesting to note that the extra numerical factor $0.1$ (between computed and observed values)  which appears in our  order of magnitude estimates  (\ref{eq:cal N ::num}) is very similar to extra factor $0.1$ which occurred in analogous  computations \cite{Zhitnitsky:2020shd} of a  number of mysterious bursts  observed by the Telescope Array and similar estimates \cite{Zhitnitsky:2021qhj} of the event rate for puzzling Multi Modal Clustering Events observed by the HORIZON 10 T instrument.   This similarity hints on a common   origin for all  three  phenomena though the  physics  for  these  phenomena  are dramatically different.  This is because  the main normalization factor representing  the   DM flux in the form of the AQN induced events   (\ref{Phi1})    is identically the same for all three  estimates, for AAE   (\ref{eq:cal N ::num}) and for Telescope Array mysterious event count \cite{Zhitnitsky:2020shd} and for Multi Modal Clustering Events  \cite{Zhitnitsky:2021qhj}.  If future studies support our identification of the AAE with the AQN induced radio pulses this numerical suppression factor might be a hint that the Standard Halo Model (which is used in  estimations for the DM flux)     underestimates the local DM density in solar system.  The true DM density locally  may dramatically deviate  from  average global value $\rho_{\rm DM}\approx 0.3{\rm\,GeV\,cm^{-3}}$, see Introduction in \cite{Liang:2020mnz} for the references and details.

\section{Radio signals induced by AQNs}
\label{sec:Radio signals induced by AQN}
 It is well known that the frequency spectrum of geosychrotron radiation by an ultrarelativistic charged particle is equivalent to that emitted by a particle moving instantaneously at constant speed on an appropriate circular path with instantaneous radius of curvature $\rho$, see  e.g. Jackson \cite{jackson}:
\begin{equation}
\label{eq:rho}
\rho
\approx\frac{\gamma mc}{e{\cal B}\sin\theta_{\cal B}}
\approx 0.8{\rm\,km}
\left(\frac{\gamma}{20}\right)
\,,
\end{equation}
where ${\cal B}\approx0.5{\rm\,gauss}$ is the local magnetic field strength, and $\theta_{\cal B}$ is the angle between the particle velocity $\mathbf{v}$ and magnetic direction. We choose $\theta_{\cal B}\approx60\degree$ in this work, as the magnetic field direction in Antarctica is approximately vertical and the exit angle of the AAEs are typically of order $30\degree$.

\begin{figure}[h]
	\centering
	\captionsetup{justification=raggedright}
	\includegraphics[width=0.8\linewidth]{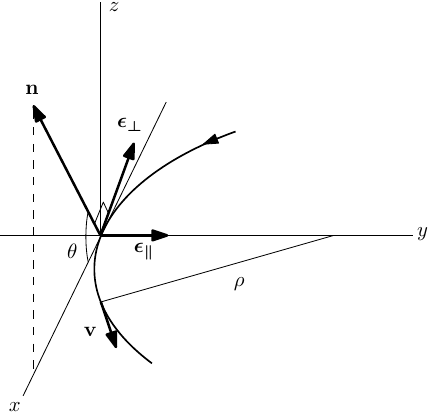}
	\caption{}.  
	\label{fig:geometry}
\end{figure}

The geometry follows from Fig. \ref{fig:geometry}, the segment of trajectories lies in the $x$-$y$ plane. The $\theta$ is the observation angle  between $\mathbf{v}$ and the direction of the observer $\mathbf{n}$. An ultrarelativisitc particle with Lorentz factor $\gamma\gg1$ has a narrow emission angle $\theta\lesssim\gamma^{-1}$, beyond which the intensity of radiation is exponentially suppressed. The $\boldsymbol{\epsilon}_\parallel$ and $\boldsymbol{\epsilon}_\perp$ are the two directions of polarization as shown on Fig. \ref{fig:geometry}.

For an observer with distance ${\cal R}$   from $N$ coherent charged particles, the spectral component of the electric field $\mathbf{E}(\omega)$ as a function of  frequency  $\omega$ is given by \cite{jackson}
\begin{equation}
\label{eqs:abs E(omega) etc.}
\begin{aligned}
&|\mathbf{E}(\omega)|
=N\left(\frac{4\pi}{c}\right)^{1/2}\frac{1}{{\cal R}}|\mathbf{A}(\omega)|\,, \\
&\mathbf{A}(\omega)
=\frac{-ie\omega}{\sqrt{8c}\,\pi}\left[
-\boldsymbol{\epsilon}_\parallel A_\parallel(\omega)
+\kappa\,\boldsymbol{\epsilon}_\perp A_\perp(\omega)
\right]\,,
\end{aligned}
\end{equation}
where $A_\parallel(\omega)$ and $A_\perp(\omega)$ corresponds to the amplitudes of two polarization directions in terms of modified Bessel functions:
\begin{subequations}
\label{eqs:A_parallel etc.}
\begin{equation}
A_\parallel(\omega)
=i\frac{2\rho}{\sqrt{3}\,c}
\left(\frac{1}{\gamma^2}+\theta^2\right)K_{2/3}(\xi)\,,
\end{equation}
\begin{equation}
A_\perp(\omega)
=\theta\frac{2\rho}{\sqrt{3}\,c}
\left(\frac{1}{\gamma^2}+\theta^2\right)^{1/2}K_{1/3}(\xi)\,,
\end{equation}
\end{subequations}
with
\begin{equation}
\label{eq:xi}
\xi
=\frac{\rho\omega}{3c}\left(\frac{1}{\gamma^2}+\theta^2\right)^{3/2}\,.
\end{equation}
Note that Eqs. \eqref{eqs:A_parallel etc.} are only well-defined for $\omega>0$, the values in the negative domain is defined by   $\mathbf{A}(-\omega)=\mathbf{A}^*(\omega)$. 
The parameter $\kappa$ in (\ref{eqs:abs E(omega) etc.}) is introduced to characterize the screening effect of $e^+e^-$ pair, where the $\boldsymbol{\epsilon}_\perp$-component is effectively cancelled ($\kappa\approx0$) when $e^+e^-$ pairs are predominantly formed in conventional CR events \cite{Huege:2003up}. In the opposite limit, we choose $\kappa\approx1$ in case of AQN-induced signal as it is primarily initiated by electrons such that the screening effect is diminished. The spectrum of electric field is therefore
\begin{equation}
\label{eq:abs E(omega)}
\begin{aligned}
|\mathbf{E(\omega)}|
&\approx\sqrt{\frac{2}{3\pi}}\frac{Ne\rho\omega}{c^2{\cal R}}
\left(\frac{1}{\gamma^2}+\theta^2\right)K_{2/3}(\xi)  \\
&\quad\times\sqrt{
1+\frac{\gamma^2\theta^2}{1+\gamma^2\theta^2}
\left(\frac{K_{1/3}(\xi)}{K_{2/3}(\xi)}\right)^2
}\,.
\end{aligned}
\end{equation}
The observational distance from ANITA balloon payload is of order 35 km and the size of the detector is of order 10 m, therefore the effective observation angle is tiny comparing to the emission angle $\theta\sim10^{-4}\ll\gamma^{-1}$. 

\begin{figure}[h]
	\centering
	\captionsetup{justification=raggedright}
	\includegraphics[width=1\linewidth]{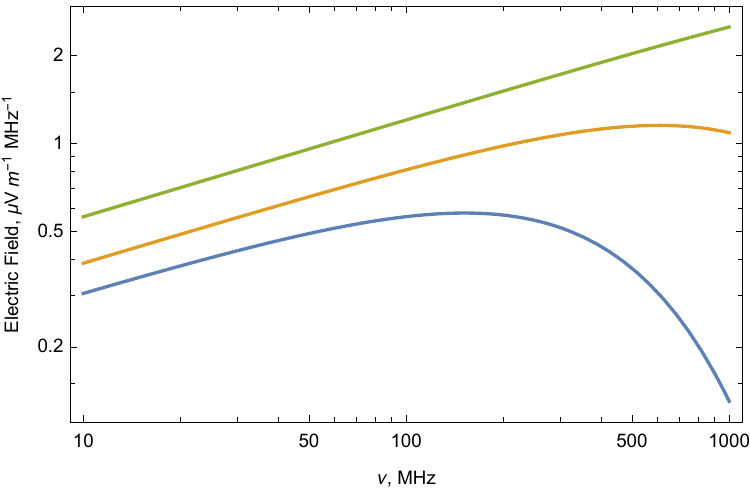}
		\caption{Spectrum of electric field $|E(2\pi\nu)|$ from Eq. \eqref{eq:abs E(omega)}, with $\theta=0$, ${\cal R}=35\,$km, and $N=5\cdot 10^8$. The Lorentz factor is chosen to be $\gamma=10$ (blue), 20 (orange), and 60 (green).}
\label{fig:E_omega}
\end{figure}

Choosing $\theta=0$, we plot the spectrum \eqref{eq:abs E(omega)} 
in Fig. \ref{fig:E_omega}  with different values of $\gamma$ for a specific value of $N=5\times 10^8$. One can see that the spectrum is very flat: the absolute value $|\mathbf{E(\omega)}|$ changes by a factor of 3 or so when the frequency $\nu$ varies by 2 orders of magnitude. Furthermore, the total strength  of the electric field integrated over entire frequency band agrees with the observed value on the level of  $|\mathbf{E}|\sim \rm mV/m$, see Refs. \cite{Gorham:2016zah,Gorham:2018ydl,Gorham:2020zne}.

Another  generic feature of synchrotron radiation is the exponential  suppression of the emission  beyond the critical frequency \cite{jackson}
\begin{equation}
\label{eq:nu_c}
\nu_{\rm c}
\equiv \frac{3\gamma^3c}{4\pi\rho}
\approx 0.7 {\rm\,GHz}
\left(\frac{\gamma}{20}\right)^2\,.
\end{equation}
This qualitative consequence of our  proposal is also consistent with
ANITA 
observations\cite{Gorham:2016zah,Gorham:2018ydl,Gorham:2020zne}.

\begin{figure}[h]
	\centering
	\captionsetup{justification=raggedright}
	\includegraphics[width=1\linewidth]{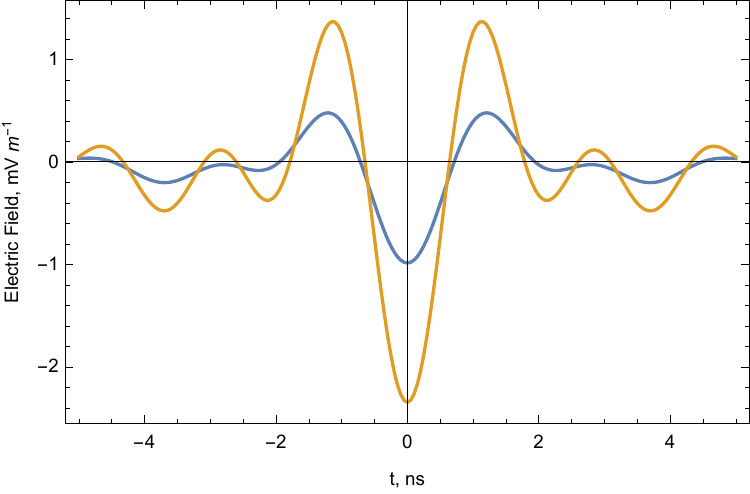}
		\caption{ Time-dependent electric field from Eq. \eqref{eq:bold E(t)} with $\theta=0$, ${\cal R}=35\,$km, and $N=5\cdot 10^8$, using an idealized rectangle filter. The Lorentz factor is chosen to be $\gamma=10$ (blue) and 20 (orange);
		filter: (40-80)\,MHz and (200-600)\,MHz.    }
	\label{fig:E_t3}
\end{figure}

The time-dependent radio pulse can be reconstructed from the frequency spectrum by an inverse Fourier transform:
\begin{equation}
\label{eq:bold E(t)}
\begin{aligned}
\mathbf{E}(t)
&=\frac{1}{\sqrt{2\pi}}\int_{-\infty}^\infty b(\omega)\mathbf{E}(\omega)e^{-i\omega t}\rmd\omega  \\
&\approx -\boldsymbol{\epsilon}_\parallel
\frac{2Ne\rho}{\sqrt{3}\pi c^2\gamma^2 {\cal R}}
{\rm\,Re}\left[
\int_0^\infty b(\omega)\,\omega {\rm K}_{2/3}(\xi)e^{-i\omega t}\rmd\omega
\right]  
\end{aligned}
\end{equation}
where $b(\omega)$ is the  filter   characterizing   the receiver,  similar to analysis in \cite{Huege:2003up}. For illustrative purpose, we plot the time-dependent electric field in Fig. \ref{fig:E_t3} by assuming an idealized rectangle filter spanning (40-80)\,MHz and (200-600)\,MHz. The pulse has an amplitude $|\mathbf{E}_0|\sim{\rm\,mV/m}$ and the time duration $\tau\sim{\rm ns}$, which is consistent with observed features of the anomalous pulses observed by ANITA \cite{Gorham:2016zah,Gorham:2018ydl,Gorham:2020zne}. Because $|\mathbf{E}(\omega)|$ is approximately flat in frequency range below $\nu_{\rm c}$, $\mathbf{E}(t)$ is essentially determined by the    inverse Fourier transform of $b(\omega)$. In case of a rectangle-like filter with frequency bandwidth $\Delta\nu$, the time duration of the pulse  is determined by $\Delta \nu$ as follows:
\begin{equation}
\label{eq:tau Delta nu}
\tau
\approx\frac{1}{\Delta\nu}
\approx 2 {\rm\,ns}\left(\frac{600{\rm\,MHz}}{\Delta\nu}\right)\,.
\end{equation}
The numerical value for  the time scale $\tau$ is not   very sensitive to the parameters of the model as explained above due to the flatness of the spectrum  below $\nu_{\rm c}$. In contrast to  the time duration   $\tau$      the absolute value of the electrical field  $|\mathbf{E}_0|\sim{\rm\,mV/m}$  is sensitive to $\gamma$ as shown on Fig. \ref{fig:E_t3}. It assumes the values  which 
are also consistent with observations   \cite{Gorham:2016zah,Gorham:2018ydl,Gorham:2020zne}.

\begin{figure}[h]
	\centering
	\captionsetup{justification=raggedright}
	\includegraphics[width=1\linewidth]{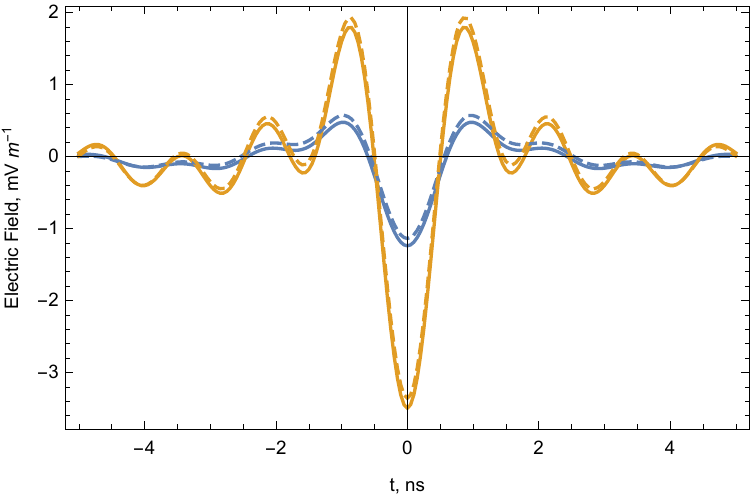}
		\caption{ Time-dependent electric field from Eq. \eqref{eq:bold E(t)} with $\theta=0$, ${\cal R}=35\,$km, and $N=5\cdot 10^8$, using an idealized rectangle filter. The Lorentz factor is chosen to be $\gamma=10$ (blue) and 20 (orange). Solid: filter spanning (40-80)\,MHz and (200-800)\,MHz; dashed: (200-800)\,MHz.  }
	\label{fig:E_t}
\end{figure}

 It is instructive to understand the temporal  features of the electric field $\mathbf{E}(t)$ by rewriting the integral (\ref{eq:bold E(t)}) in terms of the dimensionless variable $\xi$ defined by (\ref{eq:xi}) as follows:  
\begin{equation}
\label{eq:bold E(t)_temp}
\begin{aligned}
\mathbf{E}(t)
=-\boldsymbol{\epsilon}_\parallel\frac{18}{\sqrt{3}\pi \rho}
\frac{\gamma^4Ne}{{\cal R}}{\rm\,Re}\left[
\int_{\xi_{\rm min}}^{\xi_{\rm max}}\xi e^{-ia\xi}{\rm K}_{2/3}(\xi)\rmd \xi
\right]
\end{aligned}
\end{equation}
where  we assume $b(\omega)$ as an idealized rectangle filter,  $a\equiv3\gamma^3c t/\rho$ and  $(\xi_{\rm min}, \xi_{\rm max})$ are determined by corresponding values of   $\omega_{\rm min}$ and $\omega_{\rm max}$ characterizing the filter $b(\omega)$. From   (\ref{eq:bold E(t)_temp}) one can explicitly see that the typical time duration is determined by parameter  
\begin{equation}
\label{eq:tau}
 (a\xi)\approx 2\pi~~\Rightarrow ~~\tau\approx (2-4) {\rm ns},
\end{equation}
and the combination $(a\xi)$ which is the phase entering (\ref{eq:bold E(t)_temp}) is indeed $\gamma$-independent for $\theta\ll \gamma^{-1}$. 

The same formula (\ref{eq:bold E(t)_temp}) also shows that the absolute value of the field $\mathbf{E}(t)$ and time duration is mostly determined by the region of the largest values of $\xi$ close to $\xi_{\rm max}$, while the low energy portion of the spectrum does not play a role.   We explicitly checked this feature by  plotting on Fig. \ref{fig:E_t} the electric field  $\mathbf{E}(t)$ with two different models for the filter. Solid line includes both filters spanning the low   and high frequency modes, while the dashed line corresponds to a single filter describing exclusively  high frequency modes. The difference between the two curves is negligible as claimed. Another property worth to be mentioned is that the absolute value of the field  $\mathbf{E}(t)$ in the peak increases with extending the upper value for $\omega_{\rm max}$ to higher value (800 MHz on Fig. \ref{fig:E_t} versus 600 MHz on Fig  \ref{fig:E_t3}). This is also expected behaviour as the integral  (\ref{eq:bold E(t)_temp}) is saturated 
by the region of the largest values of $\xi$ close to $\xi_{\rm max}$, as already mentioned. 
   \begin{figure}[h]
	\centering
	\captionsetup{justification=raggedright}
	\includegraphics[width=1\linewidth]{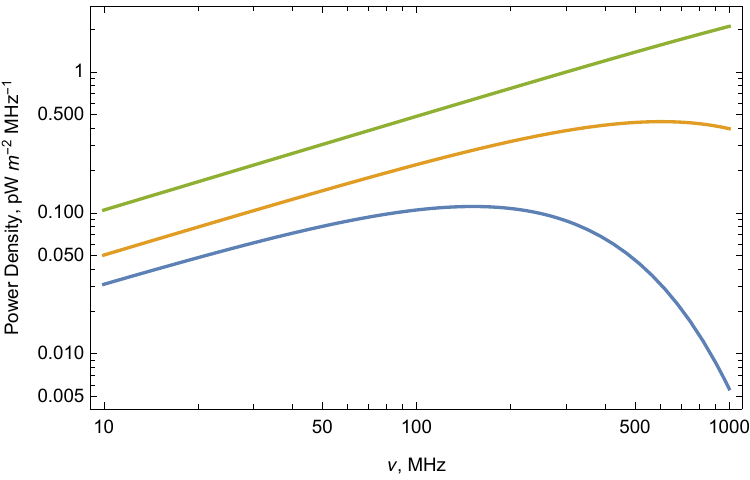}
	\caption{ Spectrum of power density from Eq. \eqref{eq:d2 P d omega dA}, with $\tau=4{\rm\,ns}$, $\theta=0$, ${\cal R}=35\,$km, and $N=5\cdot10^8$. The Lorentz factor is chosen to be $\gamma=10$ (blue), 20 (orange), and 60 (green).}.  
	\label{fig:P_omega}
\end{figure}

The AQN-induced signal is in many aspects similar to the conventional CR shower,
with crucial and dramatic  difference  being the {\it non-inverted polarity}. It well matches the observation of the anomalous events in ANITA experiment.  We list a number of distinct  features (between AQN-induced events and conventional  CR air showers events) in concluding Sec.  \ref{conclusion}.

For completeness, we also derived the spectrum of power emission similar to the estimate in Ref. \cite{Huege:2003up}. The power is related to the electric field by Poynting vector:
\begin{equation}
\mathbf{S}(t)
=\frac{c}{4\pi}\mathbf{E}(t)\times\mathbf{B}(t)
\end{equation}
where we use the Gaussian units. The power density is given by
\begin{equation}
\frac{\rmd P(t)}{\rmd A}
=|\mathbf{S}(t)|
=\frac{c}{4\pi}|\mathbf{E}(t)|^2\,.
\end{equation}
Averaging over the time duration of the pulse $\tau$, it gives
\begin{equation}
\begin{aligned}
\left<\frac{\rmd P(t)}{\rmd A}\right>_\tau
=\frac{c}{4\pi\tau}\int |\mathbf{E}(t)|^2\rmd t
\approx\frac{c}{4\pi\tau}\int|\mathbf{E}(\omega)|^2\rmd\omega\,.
\end{aligned}
\end{equation}
where the last step follows from Parseval’s theorem.   The spectrum of power density is therefore
\begin{equation}
\label{eq:d2 P d omega dA}
\frac{\rmd^2P}{\rmd \omega\rmd A}
\approx\frac{\rmd}{\rmd\omega}\left<\frac{\rmd P(t)}{\rmd A}\right>_\tau
\approx\frac{c}{4\pi\tau}|\mathbf{E(\omega)}|^2\,.
\end{equation}

The spectrum for the power of the emission shares similar properties as the one of electric field as it is expressed in terms of the electric field $|\mathbf{E(\omega)}|$.  The results for the power density is presented on Fig. \ref{fig:P_omega}, where  we use $\tau\approx 4$\,ns for numerical estimates  taken from (\ref{eq:tau}), Fig  \ref{fig:E_t3}  and  Fig. \ref{fig:E_t}. This value     is also consistent with ANITA observations shown on Fig. 2 of Ref. \cite{Gorham:2018ydl}. The power density      of order $(0.2-0.3)\rm\,pW\,m^{-2}\,MHz^{-1}$ for $\gamma=20$ and frequencies $\nu\gtrsim 100$ MHz, then it falls sharply beyond the critical frequency \eqref{eq:nu_c}. This behaviour is consistent with ANITA's results presented on Fig. 4 of Ref. \cite{Gorham:2018ydl}.
\exclude{
\begin{figure}[h]
	\centering
	\captionsetup{justification=raggedright}
	\includegraphics[width=1\linewidth]{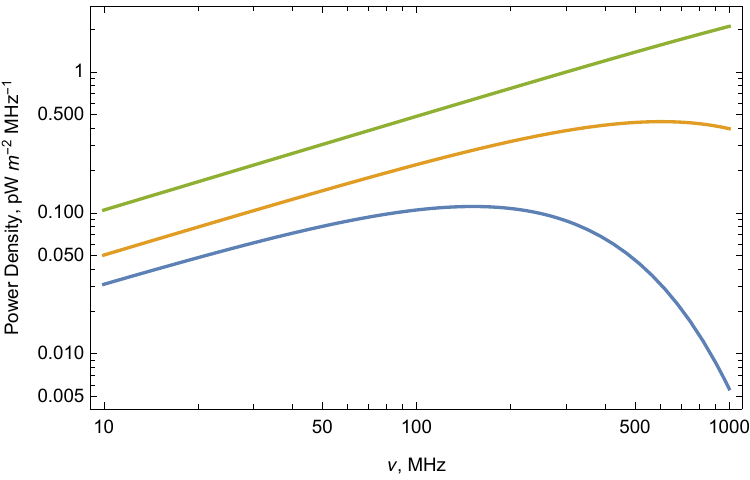}
	\caption{Spectrum of power density from Eq. \eqref{eq:d2 P d omega dA}, with $\tau=1{\rm\,ns}$, $\theta=0$, $R=35\,$km, and $N=10^8$. The Lorentz factor is chosen to be $\gamma=10$ (blue), 20 (orange), and 60 (green).}.
	\end{figure}
	}

We finish   this section with few comments on accuracy of our estimates.  It should be emphasized that the computations carried out    in this section are  oversimplified as realistic signals can be severely modified by numerous factors such as geometry of the beam, relative position of the observer, and frequency characteristic of the receiver's filters  \cite{Huege:2003up,HUEGE2005116}. Nonetheless most factors, such as lateral structure and inclined axis of the beam, are near-field effects and they are eliminated since the observation angle is small. There are many other factors, such as the electron energy distribution of the beam as a function of $\gamma$, may  modify our predictions. We cannot  predict the corresponding behaviour as it is determined by very complex non-equilibrium dynamics as discussed in Appendix \ref{details}. 

\exclude{
The most important far-field effect in our case is the longitudinal distribution of the AQN-induced electrons. For an electron beam with length $\sim c\tau$, the spectrum of the electric field involves an additional attenuation factor $S(\omega)$ \cite{Huege:2003up}:
\begin{equation}
S(\omega)
\sim\frac{1}{\sqrt{1+\omega^2\tau^2}}
\end{equation}
where we assumed the the beam follows an asymptotic $\Gamma$-distribution similar to conventional CR showers. \Red{I attached Fig. \ref{fig:E_omega_All} [including $S(\omega)$] in comparison to Fig. \ref{fig:E_omega}, which one is better?} Evidently this attenuation does not affect our conclusion as it only applies to moderate modification to the higher frequency regime $\nu\gtrsim10^2{\rm\,MHz}$. 

\begin{figure}[h]
	\centering
	\captionsetup{justification=raggedright}
	\includegraphics[width=1\linewidth]{E_omega_All}
	\caption{\Red{$\tau=1{\rm\,ns}$}}.  
	\label{fig:E_omega_All}
\end{figure}
}

\section{Conclusion and future development}\label{conclusion}
Our basic results can be summarized as follows. We explore a new possible explanation for the AAEs  \cite{Gorham:2016zah,Gorham:2018ydl,Gorham:2020zne}     by suggesting that these events can be related to the DM annihilation events within the AQN framework. To be more precise, 
we argued that these events  can be interpreted as the  AQN-induced radio pulses which  result from    the AQN traversing the Earth and going in the upward direction.
The basic qualitative characteristics (such as the emission frequency, the electric field strength and the radio pulse durations as presented in the previous section \ref{sec:Radio signals induced by AQN})  of the observed AAEs are consistent with our AQN  interpretation.

One should mention that  the AQN-induced radio  pulses   can be easily discriminated from conventional sources, such as CR air showers   \cite{Huege:2003up,HUEGE2005116}.   

Indeed, a ``rule of thumb" suggests that the maximal number of charged particles (mostly electrons and positrons) in a CR air shower is $E_{\rm CR}/{\rm GeV}$, which implies that $N\approx (10^8-10^9)$ for a $E_{\rm CR}\approx (10^{17}-10^{18})$\,eV shower (see e.g. \cite{Huege:2003up}). This number of electrons from CR shower is close  to the number of electrons being emitted by the AQN according to (\ref{N}). Furthermore, the typical average energy in the CR shower is 30\,MeV, which is also very similar to our estimates for the AQN-induced spectrum of the electrons with $E\in (1-10^2)\,$MeV with the peak around 10\,MeV according to (\ref{N}). 
Therefore, it should not be a surprise that the radio emission intensity and the electric field strength are very similar in both cases as the geomagnetic field ${\cal B}\approx0.5{\rm\,gauss}$ (which represents the source of the acceleration and consequent   radio emission)   is obviously the same in the same location.  

Now we want to discuss the drastic differences between the pulses induced by conventional CR showers and the AQNs. These dramatic distinct features can be tested in future experiments such that our proposal  can be discriminated from other suggestions. We list below   the following typical spectral features of the CR-induced radio pulses and contrast them with the AQN-induced radio pulses:

1. The generic spectral feature of the CR-induced radio emission is the presence of oscillations, which normally starts around 100 MHz (depending on the distance from the shower axis); see e.g. Fig. 1 in  \cite{HUEGE2005116}. These oscillations are due to the coherence diminishing as the wavelength becomes shorter (in comparison to the ``pancake" size in CR shower).  While it is obviously affected by the detector's filter, this feature is a physical effect due to the changing  number of coherent particles with different wavelengths. 
Such a  picture being typical for the CR-induced radio emission  is not expected to occur for the AQN-induced radio signal as the notion of a ``pancake" does not exist in our case; see also item 3 below with an argument that the notion of a ``central axis"  does not exist for the AQN-induced events.  

2. Another typical feature of the CR-induced radio emission is that the most of the power is emitted at frequencies around $(20-30)$ MHz for $E_{\rm CR}\approx 10^{17} $\,eV shower (see Fig. 2 in  \cite{HUEGE2005116}). It is a  result of very strong cutoff frequency $\nu_0\lesssim 50$ MHz, which strongly depends  on features of the shower; see Eq. (2) and (12) in  \cite{HUEGE2005116}.
It should be contrasted with our case when the cutoff frequency $\nu_{\rm c}\sim 0.7$ GHz  is determined by dramatically different physics as expressed by (\ref{eq:nu_c}). 

3. The final and the most important difference between these two cases is that cutoff frequency $\nu_0 $
in CR air showers strongly depends on many parameters of the shower, such as the distance from the central axis when number of particles per unit area strongly depends on this parameter. It must be contrasted with our case of the AQN-induced radio signal when all electrons emitted from the same point at the same instant  are moving along the same direction.  

This picture suggests that the event  could be viewed as a uniform front of size $\rho\sim 0.8$\,km defined by (\ref{eq:rho})  rather than a CR air shower with a well defined central axis.  In different words, the number of particles per unit area in the  bunch of electrons does not depend on the distance from the central axis, in huge contrast  with conventional CR air showers. The notion ``central axis" simply does not exist  for the AQN induced electrons as the number of particles per unit area is approximately the same for all electrons generating the radio pulse. 

  These  features of the AQN-induced radio events 
  are very distinct from conventional CR induced radio pulses, and it should be easily discriminated by future analyses with more quality data. It can be achieved, for example, by placing two or more independent but synchronized antennas at a distance to study the 
  same events from different locations.  It would allow to test  many ideas  advocated in the present work, including some  features of the AQN-induced radio emission which are not shared by conventional CR air showers, as discussed   in items 1- 3 above.  If future studies indeed support and substantiate    our proposal, it would be a strong argument supporting the AQN nature of the AAEs.

We conclude by mentioning that many other instruments can test the AQN model, as mentioned in sections  \ref{basics} and \ref{earth}. There are several other instruments mentioned in Refs. \cite{Sidhu:2018auv,Anchordoqui:2021xhu} can test the macroscopically large AQN-type models. However, it may require proper adjustment of bin time of detectors, as argued in Ref. \cite{Sidhu:2018auv}. We elaborate on this problem in Appendix \ref{Pierre-Auger}, given several recently published by Pierre Auger Collaboration constraints \cite{PierreAuger:2021lea,PierreAuger:2021gci} on upward-going showers. As concluded in Appendix \ref{Pierre-Auger}, the constraints \cite{PierreAuger:2021lea,PierreAuger:2021gci} rely on the assumption that the upward-going showers are seeded by primary particles with ultrarelativistic energy $E\approx (10^{16.5}-10^{18.5}) \rm\,eV$, the similar to conventional down-going showers. This assumption does not apply to our proposal because AQNs move with a non-relativistic speed $v_{\rm AQN}\sim 10^{-3} c$, so the event reconstruction so the event reconstruction qualitatively differs from those with a light speed; see     Appendix \ref{Pierre-Auger} with details.


\section*{Acknowledgements}

This research was supported in part by the Natural Sciences and Engineering
Research Council of Canada, and X.L. also by the UBC four year doctoral fellowship.

  \appendix
  \section{Some technical  details  on the AQN properties.}\label{details}
  
  In this appendix, we want to estimate the parameters which enter the formulae in the main body of the text.
  We want to estimate parameters such  as mean free path $\lambda$,  typical distance $r^*$ where $e^+e^-$ mostly produced and the electrons may leave the system, typical electric field and electric potential at distance $r^*$, etc. 
  
  
 




One should emphasize that the following evaluations will be order-of-magnitude estimations, at the very best, due to many uncertainties present in the system.   One of the  challenges is the non-perturbative QCD in the strongly coupled regime. For example, the phase diagram in this regime is not even known, as mentioned in Sec. \ref{basics}. The other challenge is the complex interaction in a non-equilibrium dynamic system. For example, a moving AQN can generate turbulence and shock waves due to its large Mach number $M=v_{\rm AQN}/c_s$. Hence, the analysis cannot be exact, even though  the AQN model has very few fundamental parameters, such as the size $R$ and the baryon charge $B$.

  

 

We start by mentioning  that pair production of $e^+e^-$ can occur in the dense and hot environment, which has been discussed in the context of quark stars at $T\gtrsim10^2\,$keV (see \cite{Usov:1997ff,Usov:2001sw, PhysRevD.70.023004,Harko_2005,Caron:2009zi,Zakharov:2010ch,Zakharov:2010yz}). 
In context of the present work all the key ingredients relevant for $e^+e^-$  production are also present in the system.  For instance, the AQN is characterized by a similar temperature $T\gtrsim 10^2\,$keV. Moreover, the quark core is assumed to be in the CS phase. Lastly, there is a strong electric field in the system. However, we cannot use the results from the previous studies obtained in the context of the quark stars. This is because the size of the AQN is much smaller than the relevant mean free paths for all elementary processes, as discussed below. As a result, thermal equilibrium cannot be achieved in the AQN system, and the non-equilibrium dynamics determine the entire physics in the high-temperature regime. It should be contrasted with  large quark stars  where the thermal equilibrium is maintained.


 
 




Nevertheless, it is instructive to review the relevant results from the previous studies \cite{Usov:1997ff,Usov:2001sw, PhysRevD.70.023004,Harko_2005,Caron:2009zi,Zakharov:2010ch,Zakharov:2010yz} on quark stars for the following reasons. First, the main ingredients in a quark star system and in our present study are similar, such as the high temperature, the dense quark matter physics, and the strong electric field. Secondly, the complexity of the computation can be demonstrated in a much simpler case, the bare quark stars where thermal equilibrium is maintained, which remains controversial 25 years after the original paper \cite{Usov:1997ff}.

It was originally suggested in \cite{Usov:1997ff,Usov:2001sw} that the quark stars can emit $e^+e^-$ pairs. Refs. \cite{Usov:1997ff,Usov:2001sw} considered a typical temperature $T\gtrsim 10^2\,$keV, which is consistent with the condition in our proposal, as mentioned in Sec. \ref{earth}. 
In Refs. \cite{PhysRevD.70.023004,Harko_2005}, the authors argued that bremsstrahlung radiation from the electrosphere could be much more important than $e^+e^-$ emission. Furthermore, the emission rate could be dramatically modified by several effects, such as the boundary effects, the inhomogeneity of the electric field, and the Landau-Pomeranchuk-Migdal suppression. In \cite{Caron:2009zi}, it was suggested that the Pauli blocking would strongly suppress the bremsstrahlung emission. More recent studies in Refs. \cite{Zakharov:2010ch,Zakharov:2010yz} argued that the so-called mean field bremsstrahlung could be the dominant mechanism.

We are not attempting to analyze all these suggested emission mechanisms critically. Instead, our goal is to mention that even a relatively simple system of the bare quark star being in equilibrium remains a matter of debate. The AQN emission at high temperatures is even more complicated due to the non-equilibrium dynamics.
 

    \exclude{
   The corresponding parameter $Q$ can be estimated as follows: 
\be
  \label{Q}
Q\approx 4\pi R^2 \int^{\infty}_{0}  n(z, T)\rmd z\sim \frac{4\pi R^2}{\sqrt{2\pi\alpha}}  \left(m T\right)   \left(\frac{T}{m}\right)^{\frac{1}{4}} , ~~
  \ee
  }
  We start by mentioning that 
  the density of positrons $n(z, T)$ in the electrosphere for the galactic environment (low temperature) 
  at distance $z$ from the nugget's surface has been computed in the mean field approximation  in \cite{Forbes:2008uf}. It    has the following form
\begin{equation}
\label{eq:nz0}
n(z, T)=\frac{T}{2\pi\alpha}\frac{1}{(z+\bar{z})^2}, 
\end{equation}
where $\bar{z}$ is the integration constant is chosen to match the Boltzmann regime at sufficiently  large $z\gg \bar{z}$. Numerical 
studies ~ \cite{Forbes:2009wg}  support the approximate analytical  expression (\ref{eq:nz0}):
\be
\label{eq:zbar}
    \bar{z}^{-1}\approx \sqrt{2\pi\alpha}\cdot m  \cdot \left(\frac{T}{m}\right)^{\frac{1}{4}}, ~ n(z=0)\approx \left(mT\right)^{\frac{3}{2}}. ~~~
\ee
  In the equilibrium  with small  annihilation rate typical for the galactic environment, the positrons will normally occupy very thin layer  on the order of $\bar{z}$ around the  AQN's quark core as computed in \cite{Forbes:2008uf,Forbes:2009wg}. However, in our case when the AQN enters the Earth's atmosphere and further the interior   a large number of non-equilibrium processes as mentioned above  are expected to occur. Furthermore,    the positron cloud is expected to expand well beyond the  thin layer  around the nugget's core
 as a result of  the  direct collisions with Earth material, in which case some positrons will be kicked off and   leave the system. 
 \exclude{
In what follows  we   assume that, to first order, the  finite portion of positrons  $\sim Q$   leave the system as a result of these complicated processes, in which case    the AQN as a system   acquires a negative  electric  charge $\sim -|e|Q$  and  get   partially  ionized,  see also Appendix \ref{details} with more details. 
}
In this case, the  one-dimensional expression (\ref{eq:nz0})   does not apply at distance   $r\gtrsim  R$ as excited positrons will be far away from the core, i.e. at distance $r\gg R$. 
   
  To proceed with our estimates we assume   that the density $ n(r, T)$ has a  power-like  behaviour at $r\gtrsim  R$ with exponent $p$.
  This assumption is consistent with our numerical studies \cite{Forbes:2009wg} of the electrosphere with $p\approx 6$. It is also consistent with conventional Thomas-Fermi model at $T=0$, see e.g. Landau textbook\footnote{In notations of Ref.
   \cite{landau-QM} the dimensionless function $\chi (x) $ behaves as $\chi\sim x^{-3}$ at large $x$. The  potential $\phi=\chi (x)/x$ behaves as 
  $\phi\sim x^{-4}$. The density of electrons in Thomas-Fermi model scales as $n\sim \phi^{3/2}\sim x^{-6}$ at large $x$.}. 
  We keep parameter $p$ to be arbitrary to demonstrate  that our main claim is not very sensitive to our assumption on numerical value of $p$.

 Therefore,   we parameterize the density as follows
\be
\label{density}
&&n(r, T)\approx n(z=0) \left(\frac{R}{r}\right)^p, ~~ R\approx 2\cdot 10^{-5} \rm cm\\ 
&&n_0\equiv n(z=0)\approx 0.16\cdot 10^{31}\left(\frac{T}{100~ \rm keV}\right)^{\frac{3}{2}}
 {\rm cm^{-3}} \nonumber 
\ee
where $ n_0\equiv n(z=0)$ is the positron  density determined by the Eq. (\ref{eq:zbar}). The density profile (\ref{density}) allows us to estimate the effective charge of the nugget $Q_{\rm eff}(r^*)$ at distance $r^*\gg R$ assuming that $Q_{\rm eff}(r^*)$ is much  greater  than the number of positrons  removed from the system\footnote{This assumption is essentially equivalent to the expectation that  the positrons removed from the system were localized at much larger distances, which is indeed the case. For example, the corresponding scale $R_{\rm cap}\approx 1$\,cm from   \cite{Zhitnitsky:2020shd} is indeed much greater than the scale $r^*\approx 10^{-3}$\,cm, which is the relevant scale of the problem to be discussed in this work.}. The corresponding  $Q_{\rm eff}(r^*)$  can be estimated by integrating from $r^*$ to infinity 
  instead of accounting for the cancellations between the original negative charge of the antimatter AQN and positive   charge of the surrounding positrons, i.e.  
 \be
 \label{Q_eff} 
   Q_{\rm eff}(r^*) &\approx& \int_{r^*}^{\infty} 4\pi r^2 n(r)\rmd r \sim \frac{4\pi n_0R^3}{(p-3)} \left(\frac{R}{r^*}\right)^{p-3}\\
    &\approx&   10^{11} \left(\frac{ 2\cdot 10^{-3} \rm cm}{r^*}\right)^{3} ~~{\rm  for}~~  p\approx 6.\nonumber
 \ee
 In estimate (\ref{Q_eff}),  we assumed that the power behaviour (\ref{density}) holds in this regime. The relevant parameter  is the charge to mass ratio $ Q_{\rm eff}/M$,  which is 14 orders of magnitude smaller for the AQNs compared to a similar ratio 
$e/m_p$ computed  for the proton. Indeed, the AQN's mass is of order $M\approx m_p B$ with a  typical baryon charge  $B\sim 10^{25}$, while $Q_{\rm eff}\sim 10^{11}$. 

Our next task is to estimate the binding energy $U(r^*)$ of the positrons at the distance $r^*$ as follows:
 \begin{equation}
 \label{U} 
   U (r^*)= \frac{\alpha Q_{\rm eff}(r^*)}{r^*}      \approx 10{\rm\,MeV} \left(\frac{2\cdot 10^{-3} \rm cm}{r^*}\right)^{4}.  
 \end{equation}
 Here we choose $r^*\approx2\cdot10^{-3}\,$cm for reasons to be explained later in this appendix.
 This estimate suggests that if an electron will be created at distance $r^*\approx 2\cdot10^{-3}$cm it will be quickly accelerated up to the energies on the order of 10 MeV as a result of strong repulsion due to the Coulomb force (\ref{U}). These parameters are used  for estimations   in the main text in Sec. \ref{proposal}. 
We shall argue below that this value $r^*\approx 2\cdot10^{-3}\,$cm is indeed an
appropriate scale, beyond which the electrophere becomes transparent to electrons created via pair production.
 
 To make our arguments more convincing, we have to estimate a number of effects related to this physics. In particular, we have to estimate  the rate 
 of $e^+e^-$ production  at this distance $r^*$, the electron's mean free path $\lambda (r^*)$, the screening length, the equilibration  time, and many other characteristics  
   which support our proposal.
   
   We start with the estimation of the $e^+e^-$ density in the environment at temperature $T$ assuming that thermal equilibrium conditions are fulfilled. The corresponding density of the electrons $n_-$ and positrons $n_+$ is given by the conventional Fermi distribution \cite{landau-statmech}: 
   \be
   \label{fermi}
   n_+=n_-=\frac{1}{\pi^2}\int_0^{\infty}\frac{p^2dp}{e^{\epsilon/T} +1}, ~~ \epsilon=\sqrt{p^2+m^2},~~~
   \ee
  where the much smaller  background  positron  density $n(r^*)$ from (\ref{density}) has been ignored in the estimates (\ref{fermi}). 
At very low temperatures $T$, the number densities are suppressed by a factor of $\exp (-m/T)$. This is because the number density of energetic photons capable of producing massive particles is exponentially small.
   
   
 




Formula \eqref{fermi} is valid only in the large volume   limit so that the thermal equilibrium can be maintained. By ``large volume'', we require all mean free paths for all processes are much smaller than the size of the system $V$. However, this condition is not satisfied in our case of small nugget, as we will argue below.





 
  In particular, the typical cross sections for $ \gamma \gamma\rightarrow e^+e^-$, $ e^+e^- \rightarrow  \gamma \gamma $, and $\gamma e^{\pm} \rightarrow \gamma e^{\pm}$ are on the order of
  \begin{equation}
  \label{cross-section}
  \begin{aligned}
    & \sigma_{\gamma \gamma\rightarrow e^+e^-}\sim \sigma_{e^+e^- \rightarrow  \gamma \gamma}\sim \sigma_{\gamma e^{\pm} \rightarrow 
    \gamma e^{\pm}}  \sim  \pi r_0^2\,, \\
    &r_0=\frac{\alpha}{m}\approx 2.8\cdot 10^{-13} \rm\,cm\,. 
  \end{aligned}
  \end{equation}
  at relativistic  velocities and temperature $T\approx 100 $\,keV.
   The mean free path for these processes can be estimated as follows  
   \be
   \label{lambda}
   \lambda\sim \frac{1}{\sigma n_{\pm}}\sim 2.5\cdot 10^{-3} \rm cm, 
   \ee
   which is  the same order of magnitude as  the  $r^*$ entering Eqs.  (\ref{Q_eff}) and (\ref{U}). This observation  obviously implies that the thermal equilibrium cannot be maintained in  a small volume   $V\sim (r^*)^3$ if $r^*\ll   \lambda$. 
   
   Another process that equilibrates the system  is the Coulomb elastic scattering with cross section $\sigma_{\rm Coul}$   estimated  as follows: 
    \begin{equation}
    \label{Coulomb}
    \sigma_{\rm Coul}\approx \frac{\alpha^2}{E^2\theta^4} \approx 0.8\cdot 10^{-25} \left(\frac{1}{\theta}\right)^4 \left(\frac{m}{E}\right)^2\rm cm^2\,,
        \end{equation}
     which is essentially the same order of magnitude as (\ref{cross-section}) for $\theta\sim 1$. Furthermore, the processes related to $ \sigma_{\rm Coul}$, $\sigma_{e^+e^- \rightarrow  \gamma \gamma}$, and $\sigma_{\gamma e^{\pm} \rightarrow 
    \gamma e^{\pm}}$ decrease   at large energies as $E^{-2}$ as explicitly shown in (\ref{Coulomb}), which makes typical  $\lambda$ estimated in (\ref{lambda})  much greater  at larger energies:   
    \be
   \label{lambda1}
   \lambda (E)\sim    \frac{1}{\sigma (E)n_{\pm}}\sim 8\cdot 10^{-3}  \left(\frac{E}{m}\right)^2 \rm cm.
   \ee
    
    We mention dependence $\lambda (E)$ on energy $E$ because the key element of our proposal is the fast acceleration of the produced electrons due to the strong background repulsive (for electrons) electric field determined by (\ref{U}).     
    Our estimates above strongly suggest that the electrons produced by this mechanism can get accelerated up to 10\,MeV energy without much scattering. This is because $\lambda(E)$ is much larger than the size $r^*\sim10^{-3}\,$cm of the accelerating region.

      


Similar arguments also suggest that there is a suppression factor $ ({r^*}/{\lambda})^3 \ll 1$ if $\lambda\gg r^*$. This factor accounts for the ``reduced'' equilibration volume, violating the basic requirement for maintaining the thermodynamical equilibrium for $\lambda\gg r^*$. 

To make this argument more quantitative, we estimate $r^*$ from the requirement $\lambda\approx r^*$, which determines the numerical value of  $r^*\approx 2\cdot 10^{-3} \rm cm$ entering Eqs. \eqref{Q_eff} and \eqref{U}.

      
      \exclude{we define $\lambda(r^*)$ as the mean free path computed with ``reduced" effective density $  n_{\pm}({r^*}/{\lambda})^3$. The requirement $\lambda(r^*)\sim r^*$ defines an approximate numerical value for $r^*$ when produced electrons 
      can get accelerated without much scattering on the way as the effective mean free path $\lambda(r^*)$ is the same order of magnitude as the region of acceleration $\sim r^*$. 
      }
     
     There is an additional suppression related to dramatically different time scales: the typical time for the equilibration is $\lambda/c$ is much longer than the typical time of electron's acceleration to relativistic velocity which is $t_0\sim m r^*/ U(r^*)$. After this short period of time $t_0$ the accelerated electrons  are lost for the   equilibration with other particles. Accounting for both these effects 
     the maximal number of $e^+e^-$ pairs being produced by this mechanism in volume $V\sim (r^*)^3$ can be estimated as follows:   
\beq
  \label{N1}
  \left[n_{\pm} \cdot V \right]\cdot \left(\frac{r^*}{\lambda}\right)^3\cdot\left(\frac{t_0 c}{\lambda}\right) \sim 10^{15}\,,\quad V\sim (r^*)^3\,.
  \eeq 
   Another possible suppression factor could be due to a strong overestimation of  the effective volume in  \eqref{N1} approximated  as $V\sim (r^*)^3\sim \lambda^3$.
   The point is that the $\gamma$ radiation and accompanied pair production of  $e^+e^-$ from this region become the dominant cooling mechanism
   only in the vicinity where the quark-antiquark pair gets annihilated within a small area $\pi R^2$. The equilibration time $\lambda/c$ is too long to distribute this heat over the entire volume $V\sim (r^*)^3$ sufficiently quickly before the particles leave the system.  Therefore, we expect an additional 
    suppression factor in  \eqref{N1} which accounts for this effect: 
\begin{equation}
\label{eq:V_eff V}
 \left(\frac{R}{r^*}\right)^2
\sim10^{-4}\,.
\end{equation}
Accounting for this additional suppression further reduces the estimate (\ref{N1}) for the maximal number of available electrons: 
\be
  \label{N}
  N_{\rm max}\sim \left[n_{\pm} \cdot V\right]  \left(\frac{r^*}{\lambda}\right)^3 \left(\frac{t_0 c}{\lambda}\right)   \left(\frac{R}{r^*}\right)^2 \sim 10^{11}.~~~~
  \ee  

 In addition to these suppression factors, there is a  number of many-body effects which may modify the estimate (\ref{N}).   For example,    there  is the Debye screening with the  corresponding length $ \lambda_D$ being defined  by the  formula
  \be
  \label{Debye}
  \lambda_D\approx \sqrt{\frac{T}{4\pi\alpha n (r, T)}}~~,
  \ee
    where $n (r, T)$ is the background positron density determined by (\ref{density}). 
    
    The Debye screening  normally applies to the situation when 
    a single external charge $q$ is inserted into the plasma. In this case, the charge $q$  will be screened on the scale of  $\lambda_D$, i.e $q\exp
    (- {r^*}/{\lambda_D}) $.  There are a few assumptions for the 
    Debye screening to be operational. First of all, the elementary processes responsible for the screening should be  much faster than the typical time scales of a slowly moving   external charge $q$. Secondly, the density of the external charges $q$ must be much smaller than the density of the charged particles  in the surrounding plasma. Both assumptions are not justified   in our case when we treat the produced electrons as the inserted external charges.  
    
    Indeed, the mean free path of the positrons from electrosphere, which are capable to screen the emitted electrons,  is $\lambda^{-1}(r^*)\sim \sigma n (r^*, T)$, where  $\sigma$ is defined by (\ref{cross-section}) and $n (r^*, T)$ is defined by (\ref{density}). The numerical value for $\lambda (r^*)$   is  much greater  than $\lambda_D$, i.e.
 $\lambda (r^*) \gg  \lambda_D$. Therefore, the effective  screening occurs on scales of order of  $\lambda (r^*)$, not $\lambda_D$.
    
    Furthermore, the density (\ref{fermi}) of the $e^+e^-$ pairs (which are treated as external charges) is also greater than background density (\ref{density}) at distance $r^*$.  Therefore, we do not expect that 
     the Debye screening is operational in the present circumstances, and we neglected it in our order of magnitude estimate (\ref{N}).  
    \exclude{Nevertheless, if we 
    literally apply 
 the Debye screening to the produced electrons (\ref{N}) we will have a suppression factor on the level of 
 \be
 \label{suppression}
 \exp{\left(-\frac{r^*}{\lambda_D} \right)} \sim 10^{-5}.
 \ee
 }
 
One should also mention that the density of photons $n_{\gamma}$ at this temperature $T$ in this region is much higher than the  density of $e^+e^-$ pairs (\ref{fermi}) by at least a factor of $ ({2\pi}{\alpha}^{-1})\exp(2m/T)$ (see Appendix B  in \cite{Zhitnitsky:2021qhj}). These photons with typical frequencies  $\omega\sim T$  may leave the system. In fact, they provide the dominant cooling mechanism in the atmosphere, as we already mentioned in the main text. However, we are not interested in the fate of these photons in the present work, as they will be quickly absorbed within relatively short distances from the path of the AQN.

 
 
 
The arguments above conclude numerous suppression factors entering \eqref{N}. It implies a considerable uncertainty in the estimates for the effective number $N$ of the emitted electrons, which are accelerated to 10\,MeV energies and leave the system. The emitted electrons can propagate for up to several kilometres, the mean free path in the atmosphere for such energetic electrons. With such uncertainty, we treat $N$ as a parameter constrained by $N\ll N_{\rm max}$ in the main text. A precise estimate is challenging as it is a prerogative of the non-equilibrium dynamics, a topic well beyond the scope of the present work.
 
 
 \exclude{
 One may wonder why we choose the key scale of the problem to be $r^*\approx 10^{-3}$\,cm and not much smaller or much greater than this scale. The answer is as follows. If we take much smaller scale, let us say one order of magnitude smaller than $r^*\approx 10^{-3}$\,cm, the typical number of particles $N_{\rm max}$ would be 4 orders of magnitude fewer than the estimate (\ref{N}). On other hand if we take much greater scale, let us say few times  larger  than $r^*\approx 10^{-3}$\,cm we would get electric field   (\ref{U}) to be too weak to accelerate the particles to  sufficiently high energies  when they can leave the system without much re-scattering and eventual annihilation with surrounding positrons.
 }
 In our estimates, we assume that the most of the electrons will be accelerated from the region $\sim r^*$ because from this region the electrons may 
 get accelerated without much collisions with positrons from the electrosphere. 
However, we generally expect that the accelerated particles cover the entire range of the spectrum, $E\in(1-10^2)\,$MeV, which can leave the system. For qualitative analysis, we assume that the peak of this distribution is on the order of 10\,MeV, i.e.
 \begin{equation} 
 \label{E}
\la E\ra \sim U(r^*) \sim 10{\rm \,MeV}\,,\quad
E\in (1-10^2){\rm\,MeV}\,,
\end{equation}
which is precisely the magnitude we used in all our estimates in the main text.


 \section{Pierre Auger  constraint on upward-going showers }\label{Pierre-Auger}
It has been recently published in   \cite{PierreAuger:2021lea,PierreAuger:2021gci} the constraints on upward-going showers with the Fluorescence Detector (FD). The upper limits for upward-going showers with energies larger than $E> 10^{17.5} \rm eV$ have been formulated as 
\be
\label{AUGER-constraints}
F_{\gamma=1}^{95\%}=\rm \frac{3.6\cdot 10^{-20}}{cm^2 \cdot s\cdot sr}\simeq \frac{1.1\cdot 10^{-2}}{km^2 \cdot yr\cdot sr}\nonumber \\
F_{\gamma=2}^{95\%}=\rm \frac{8.5\cdot 10^{-20}}{cm^2 \cdot s\cdot sr}\simeq \frac{2.6\cdot 10^{-2}}{km^2 \cdot yr\cdot sr},
\ee
 where  parameter  $\gamma$ is defined as the spectral index $E^{-\gamma}$ for  upward-going showers induced by primary particle with energy $E$. 
 We also expressed the upper limits in the same units $\rm km^{-2}\cdot yr^{-1}$ which we used for the AQN flux hitting the Earth as defined by (\ref{eq:cal N}), (\ref{Phi1}) for comparison. 
 
 One can naively think that the upper limit constraints (\ref{AUGER-constraints}) are already  in contradiction with the AQN flux (\ref{Phi1}) which has been used in our estimates. However, in fact there is no contradiction here as the upper limit constraints (\ref{AUGER-constraints}) cannot be directly applied to the AQN upward events. The main point here is that the AQNs are slow moving     non-relativistic objects with typical velocity $v_{\rm AQN}\sim 10^{-3}c$ while the conventional analysis assumes that the primary particle is  ultra-relativistic particle moving with the speed of light $c$. It obviously  implies that all bin time detectors are not properly adjusted for such slow moving AQNs. 
 
 Furthermore, the discrimination of the signal from background events is achieved    by using the Profile Constrained Geometry Fit Reconstruction (PCGF) that forces the depth profile to match the approximately universal characteristics of air showers induced by most primaries with well defined   shower maximum   of known width. As we discussed in length in this work such notions as  a ``central axis" and a ``pancake" do not exist in the AQN framework. Instead the AQN induced  
  event  should  be viewed as a uniform front when the number of particles per unit area does not much depend on the distance from the 
  central axis\footnote{This last feature plays a very important role in our prediction for the radio-signal induced by the AQN, as remarked  in Conclusion in item 1.
  }.
  It is obvious that this picture of the AQN-induced event is dramatically different from the   criteria being used in \cite{PierreAuger:2021lea}  to select  a proper upward-going air shower 
  reconstructed by PCGF.  
  
  Therefore, the upper limits (\ref{AUGER-constraints}) do not impose any constraints on the AQN-induced events.  At the same time, the upper limits (\ref{AUGER-constraints}) strongly constraint the physical models  when the shower  is initiated by $\tau$ leptons being produced from $\tau$-neutrinos  (or some BSM particle) interacting with the Earth's material \cite{PierreAuger:2021gci} because in this scenario the shower is expected to have the conventional universal characteristics
 such that the PCGF reconstruction   properly selects such events. 
 
\exclude{
\newpage\section{\Red{temporary section, plots comparison}}
 
\begin{figure}[htp!]
	\centering
	\captionsetup{justification=raggedright}
	\includegraphics[width=1\linewidth]{E_omega}
	\caption{Spectrum of electric field $|E(2\pi\nu)|$ from Eq. \eqref{eq:abs E(omega)}, with $\theta=0$, $R=35\,$km, and $N=10^8$. The Lorentz factor is chosen to be $\gamma=10$ (blue), 20 (orange), and 60 (green).}
\end{figure}

\begin{figure}[htp!]
	\centering
	\captionsetup{justification=raggedright}
	\includegraphics[width=1\linewidth]{E_omega_N9}
	\caption{\Red{$N=10^9$, Spectrum of electric field}}
\end{figure}

\begin{figure}[htp!]
	\centering
	\captionsetup{justification=raggedright}
	\includegraphics[width=1\linewidth]{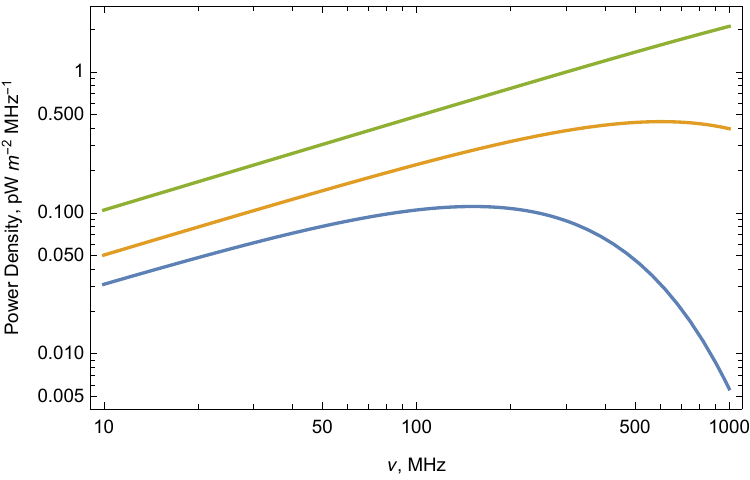}
	\caption{Spectrum of power density from Eq. \eqref{eq:d2 P d omega dA}, with $\tau=1{\rm\,ns}$, $\theta=0$, $R=35\,$km, and $N=10^8$. The Lorentz factor is chosen to be $\gamma=10$ (blue), 20 (orange), and 60 (green).}.  
\end{figure}
\begin{figure}[htp!]
	\centering
	\captionsetup{justification=raggedright}
	\includegraphics[width=1\linewidth]{P_omega_N58}
	\caption{\Red{$N=5\times10^8$, Spectrum of power density}}
\end{figure}
\begin{figure}[htp!]
	\centering
	\captionsetup{justification=raggedright}
	\includegraphics[width=1\linewidth]{P_omega_N9}
	\caption{\Red{$N=10^9$, Spectrum of power density}}
\end{figure}
\newpage

}
 
\bibliography{ANITA}

\end{document}